\documentstyle[12pt]{article}
\def\d{\delta}

\def\dev{\partial}
\def\G{\Gamma}
\def\m{\mu}
\def\n{\nu}

\def\l{\lambda}
\def\T{\Theta}

\font\tenrm=cmr12
\font\tenit=cmti12

\def\pmb#1{\setbox0=\hbox{#1}%
  \kern-.025em\copy0\kern-\wd0
  \kern.05em\copy0\kern-\wd0
  \kern-.025em\raise.0433em\box0}

\def\bfxi{\pmb{$\xi$}}
\begin{document}
\begin{titlepage}
\begin{flushright}
IFUP-TH-43/93
\end{flushright}
\vskip .9truecm
\begin{center}
\LARGE \bf
2+1 GRAVITY AND CLOSED TIME-LIKE CURVES\footnote{Work
partially supported by M.U.R.S.T}
\end{center}
\vskip 1truecm
\begin{center}
{{\tenrm P. Menotti} \\ 
{\tenit Dipartimento di Fisica dell' Universit\`a, Pisa 56100,
Italy and}\\ 
{\tenit INFN, Sezione di Pisa}\\
{\tenrm and}\\
{\tenrm D. Seminara} \\ 
{\tenit Scuola Normale Superiore, Pisa 56100, Italy and}\\ 
{\tenit INFN, Sezione di Pisa}}
\end{center}
%
%
\vskip .8truecm
\begin{center}
Talk given at the Workshop on ``Constraint Theory and Quantization
Methods'' \\
Montepulciano, Italy, June 1993
\end{center}
\end{titlepage}
\begin{abstract}
\noindent
In this paper we report some results obtained by applying the radial
gauge to 2+1 dimensional gravity.  The general features of this gauge
are  reviewed  and it is shown how they allow the general solution of
the problem
in terms of simple quadratures. Then we concentrate on the general
stationary problem providing the explicit solving formulas for the metric
and the explicit support conditions for the energy momentum tensor.
The chosen gauge allows, due to its physical nature,
to exploit the weak energy condition and in this connection it is
proved that for an open universe conical at space infinity the weak
energy condition and the
absence of closed time like curves (CTC) at space infinity imply the
total absence of CTC. It is pointed out how the approach can be used to examine
cosmological solution in 2+1 dimensions.
\end{abstract}

\section{ Introduction}

Gravity in 2+1 dimensions \cite{flat} turned out to be a good theoretical
laboratory both at the
classical and at the quantum level. In addition to be interesting in itself,
the theory is important in connections to the cosmic
strings \cite{string},
as all solutions in 2+1 dimensions are special solution of 3+1
dimensional gravity.

Most attention has been devoted in the past to point like or string like
sources and to stationary problems, even though some inroads \cite{MS2,hooft}
 have been made
in the realm of the time dependent problem.

It has been shown \cite{MS2,MS1,MS3,MS6,MS7}
that a special choice of gauge allows to
give general resolvent formulas for the metric in terms of simple quadratures
both in the case of time dependent and extended sources. The main reason is
the practical identification in 2+1 dimensions of the Riemann and
Ricci tensors which
allows to reformulate the problem as the solution of the covariant conservation
and symmetry constraints on the energy momentum tensor.
The procedure of solution is such that one has a complete control on the
support properties of the energy momentum tensor; still more important
is the fact that
due to the physical nature of the gauge, one is able to exploit the weak
energy  condition (WEC)  without the imposition of which, Einstein's equations
loose most of their content.
The possibility of exploiting  the WEC will be instrumental in
the problem of the occurrence of closed time-like curves (CTC).

In this paper we shall give a brief survey of the techniques and the
results obtained by exploiting the radial gauge, referring for details to ref.
\cite{MS2,MS1,MS3,MS6,MS7}.

As in connection with the problem of CTC we shall be mainly interested in the
stationary case, we shall report and discuss in sec.2 in more detail the
resolvent formulae and the support conditions for the stationary case, which
will be dealt with by developing a variant of the general radial gauge, i.e.
the reduced radial gauge which is more apt to the time independent situation.

Turning to  the  problem of CTC  \cite{CTC},
 in sect.3 we shall prove the following
result \cite{MS3,MS6}:
for a stationary solution with rotational symmetry the imposition of
i) the weak energy condition (WEC) and ii) the absence of CTC at space infinity
prevents the occurrence of CTC everywhere in an open (conical) universe.

An extension is  given of the same result to any stationary solution, also
in absence of rotational symmetry, provided that in our coordinate system the
determinant of the dreibein never vanishes \cite{MS7}.

\section{ General solution in the radial gauge}

The radial gauge, which can be defined in any space-time dimensions \cite{MT},
presents particular features in 2+1 dimensions due to the practical
identification of the Riemann and Ricci tensors. The defining equations
are
\begin{equation}
\xi^\mu\G^a_{b\mu}=0,
\end{equation}
\begin{equation}
\xi^\mu e^a_\mu=\delta^a_\mu\xi^\mu.
\end{equation}
These conditions define the usual Riemann-normal coordinates on the
manifold and in this gauge one can express the connection and the vierbein
in terms of the  Riemann two-form as follows
\begin{equation}
\label{radsol1}
\Gamma^{a}_{b\mu}(\xi)=\xi^\rho\int^1_0
R^{a}_{b\rho\mu}(\lambda\xi)\l d\l
\end{equation}
\begin{equation}
\label{radsol2}
e^a_\mu(\xi)=\d^a_\m +\xi^\rho\xi^b\int^1_0 R^a_{b\rho\mu}(\l\xi)\l(1-\l)d\l.
\end{equation}
As in 2+1 dimensions the Riemann two-form through Einstein's equations is
directly given in term of the energy momentum form Eqs. \ref{radsol1}
and \ref{radsol2} express the geometry of the space in term of the sources
through a simple
quadrature. On the other hand the energy momentum form is not arbitrary
but it is subject to the symmetry and covariant conservation
conditions which are nothing
else than Bianchi identities. Thus in the present approach the problem is
reduced to constructing the most general energy momentum form which is
symmetric and covariantly conserved. The solution of such  constraints
can be given  through a simple quadrature \cite{MS2}.
For the stationary problem, with which we shall be mainly concerned
here, the radial gauge as formulated above is in general not apt
due to the fact that it singles out a special event in space time.
One can however, for stationary problems, define a similar gauge
which we shall call reduced radial gauge, through the conditions
(in the following $i,\ j,\ l$ run over space indices)
\begin{equation}
\xi^i\Gamma^a_{bi}(\bfxi)=0
\end{equation}
\begin{equation}
\xi^i e^a_{i}(\bfxi)=\xi^i\delta^a_i.
\end{equation}
This gauge has a natural interpretation as the reference frame of an
observer which follows an integral curve of the (time-like) Killing field.
It corresponds to the Fermi-Walker  coordinates \cite{MS1,MS6,Fermi}.
The resolving
formulae analogous to Eqs. \ref{radsol1} and \ref{radsol2} are
\begin{equation}
\Gamma^a_{bi}({\bfxi})=\xi^j \int^1_0 R^a_{bji}(\l{\bfxi})\l d\l,
\end{equation}
\begin{equation}
\Gamma^a_{b0}({\bfxi})=\G^a_{b0}({\bf 0})+\xi^i\int^1_0 R^a_{bi0}(\l {\bfxi})
d\l,
\end{equation}
\begin{equation}
\label{eai}
e^a_i=\delta^a_i+\xi^j\xi^l\int^1_0 R^a_{jli}(\l{\bfxi})\l(1-\l) d\l,
\end{equation}
\begin{equation}
\label{ea0}
e^a_0=\delta^a_0+\xi^i\G^a_{i0}({\bf 0})+\xi^i\xi^j\int^1_0 R^a_{ij0}
(\l{\bfxi})(1-\l) d\l.
\end{equation}
In 2+1 dimensions the Riemann two-form appearing in the previous equations
is given  in terms of the energy momentum form $T_c$ by
\begin{equation}
\label{einstein}
\varepsilon_{abc} R^{ab}= -{2\kappa} T_c,
\end{equation}
where $\kappa=8\pi G$, and thus
\begin{equation}
R^{ab}=-{\kappa}\varepsilon^{abc} T_c=-{\kappa\over 2}
\varepsilon^{abc}~\varepsilon_{\rho\m\n}\tau^{~\rho}_c dx^\m\wedge dx^\n.
\end{equation}
Using such a relation one can
express through  a simple quadrature, the connections and  the
vierbeins in terms of the energy momentum tensor, which is the source of
the gravitational  field and thus one solves Einstein's equation.
We come now to the covariant conservation and symmetry constraints on the
energy momentum tensor. The  problem is to construct
the general
conserved symmetric energy momentum tensor
in the reduced radial gauge,  which in addition
should satisfy other physical requirements given by the support of the
sources and the restrictions due to  the energy condition \cite{H.E.}.

The conservation and symmetry equations for the energy momentum tensor are
\begin{equation}
\label{conservation}
{\cal D} T^a=0,
\end{equation}
\begin{equation}
\label{symmetry}
\varepsilon_{abc} T^b\wedge e^c =0.
\end{equation}
The most general solution of Eq. \ref{conservation} is \cite{MS6}
\begin{eqnarray}
\label{e.m.t.}
\tau^\rho_c(\bfxi)&&=\frac{1}{\kappa} \biggl [ P^\mu \partial_\mu A^\rho_c
(\bfxi)
-\frac{1}{\rho} A^\rho_c (\bfxi)-\frac{1}{\rho} \Theta^\rho \Theta_\mu
A^\mu_c(\bfxi)- P^\rho \biggl (\partial_\mu  A^\mu_c(\bfxi)-\nonumber\\
&&-\frac{1}{2}\varepsilon_{clm}\varepsilon_{\alpha\beta\sigma} P^\alpha
 A^{l\beta}(\bfxi) A^{m\sigma}(\bfxi)\biggr )\biggr ],
\end{eqnarray}
where $A^\mu_c$ is an arbitrary field. The field $A^\mu_c$ is related to
the connection $\Gamma^{ab}_\mu$ in the reduced radial gauge by
\begin{equation}
\label{connection}
\Gamma^{ab}_\mu(\bfxi)=\varepsilon^{abc}\varepsilon_{\mu\rho\nu}P^\rho
A^\nu_c(\bfxi).
\end{equation}
More demanding is the imposition of the symmetry property Eq.
\ref{symmetry} which however can be solved as follows. One express
$A^\mu_c(\bfxi)$ in component form
\begin{equation}
\label{A}
A^\rho_c(\bfxi)= T_c \left [ \Theta^\rho\beta_1+T^\rho\frac{(\beta_2-1)}{\rho}
\right ]+\Theta_c \left [\Theta^\rho\alpha_1 +T^\rho\frac{\alpha_2}{\rho}
\right ]+P_c \left [\Theta^\rho \gamma_1+T^\rho\frac{\gamma_2}{\rho}
\right ].
\end{equation}
where
$\displaystyle{T_{\mu}=\frac{\partial \xi^0}{\partial \xi^\mu}}$,
$\displaystyle{P_{\mu}=\frac{\partial \rho}{\partial \xi^\mu}}$ and
$\displaystyle{\Theta_{\mu}=\rho\frac{\partial \theta}{\partial \xi^\mu}}$
are the cotangent vectors
defined by  the polar variables in the $(\xi^1,\xi^2)$ plane.
This gives the following expression for $\tau^{a\rho}$
\begin{eqnarray}
\label{taumisto}
&&\tau^{\rho}_c=-\frac{1}{\kappa}\biggl\{ T_c \biggl ( T^\rho
\frac{\beta^\prime_2}{\rho}+\T^\rho \beta^\prime_1\biggr )+
\T_c
\biggl ( T^\rho
\frac{\alpha^\prime_2}{\rho}+\T^\rho\alpha^\prime_1
\biggr )+\nonumber\\
&&P_c\biggl (T^\rho \frac{\gamma^\prime_2}{\rho}+
\T^\rho
\gamma^\prime_1\biggr ) +
\frac{1}{\rho}P^\rho
\biggl [ T_c\biggl (\alpha_1\gamma_2-\alpha_2\gamma_1-
\frac{\dev\beta_1}{\dev \theta}\biggr )+\nonumber\\
&&\T_c \biggl (\beta_1\gamma_2-\beta_2\gamma_1-
\frac{\dev\alpha_1}{\dev \theta}\biggr )
+P_c \biggl (\alpha_1\beta_2
-\alpha_2\beta_1-
\frac{\dev\gamma_1}{\dev \theta}\biggr )\biggr ]\biggr \}.
\end{eqnarray}
Introducing the  primitives
of the functions $\alpha_1,\ \beta_1,\ \alpha_2,\ \beta_2$
\begin{eqnarray}
&&A_1(\bfxi)=\rho \int^1_0 \alpha_1(\lambda \bfxi) d\lambda-1\ \ {\  ,\  }
\ \ \ B_1(\bfxi)=\rho \int^1_0 \beta_1(\lambda \bfxi) d\lambda \nonumber,   \\
&&A_2(\bfxi)=\rho \int^1_0 \alpha_2(\lambda \bfxi) d\lambda\phantom{-1}
\ \ {\rm and}
\ \ \ B_2(\bfxi)=\rho \int^1_0 \beta_2(\lambda \bfxi) d\lambda,
\end{eqnarray}
the symmetry
condition is reduced to the following system of differential equations
\begin{eqnarray}
\label{22a}
&&A_1\alpha_2-A_2 \alpha_1+B_2
\beta_1-B_1\beta_2=0\\
\label{22b}
&&A_2\gamma_1 -A_1
\gamma_2 +\frac{\partial B_1}{\partial\theta}=0\\
\label{22c}
&&B_2\gamma_1-B_1
\gamma_2 +\frac{\partial A_1}{\partial\theta}=0.
\end{eqnarray}
In general, in absence of rotational symmetry, caustics may develop in
the sense that geodesics emerging from the origin with different
$\theta$ can intersect at some point for large enough $\rho$. This renders
the map of $\rho,\theta$ into the physical points of space not one to
one, but the geometry can be still regular in the sense that a proper
change of coordinates removes the singularity. For an example of how
this non single valuedness can show up and how it can be removed by changing
coordinates, we
refer to the appendix  of ref. \cite{MS1}. Such a
problem does not arise in the case of rotational symmetry.

We recall furthermore that to give a
regular geometry,  the functions $\alpha_i,\ \beta_i,\ \gamma_i$ must satisfy
simple regularity conditions at the origin \cite{MS2,MS6}.

Eqs. \ref{22a}, \ref{22b}, \ref{22c}
give  the whole geometry of the problem once three of
the functions, e.g. $\alpha_1,\ \beta_1,\ \gamma_1$, are given as data; in
fact  the other three can be obtained by  a single quadrature \cite{MS6}.
We have
\begin{equation}
\label{a2}
\alpha_2=\frac{B^2_1}{B_1^2-A^2_1}
\frac{\partial }{\partial\rho}\left (\frac{N}{ B_1}\right )+
2\alpha_1 I.
\end{equation}
\begin{equation}
\label{b2}
\beta_2=\frac{A^2_1}{B_1^2-A^2_1}
\frac{\partial }{\partial\rho}\left (\frac{N}{ A_1}\right )+
2\beta_1 I,
\end{equation}
\begin{equation}
\label{g2}
\gamma_2=\frac{B^2_1}{B_1^2-A^2_1}
\frac{\partial }{\partial\theta}\left (\frac{A_1}{ B_1}\right )+
2\gamma_1 I,
\end{equation}
where
\begin{equation}
\label{N}
N(\rho,\theta)
\equiv A_2 B_1 -A_1 B_2=\frac{1}{2\gamma_1}\frac{\partial}{\partial\theta}
(A_1^2-B_1^2)
\end{equation}
and coincides with the determinant of the dreibein in polar coordinates, while
$I$ is given by
\begin{equation}
\label{I}
I=\int^\rho_0 d\rho^\prime \frac{N (A_1\beta_1-B_1 \alpha_1)}{(B_1^2-
A^2_1)^2}.
\end{equation}
This parametrization of the source allows  a simple characterization
 of the  support properties  of the energy momentum tensor. In fact
one can prove \cite{MS6} that if the energy momentum tensor vanishes for $
\rho>\rho_0(\theta)$ one has
\begin{equation}
\label{support1}
\alpha_1 B_1- A_1 \beta_1={\rm constant}\ \ \ {\rm  for}
\ \ \rho>\rho_0(\theta)
\end{equation}
\noindent
and
\begin{equation}
\label{support2}
\alpha^2_1 - \beta^2_1 + \gamma^2_1={\rm constant}\ \ \ {\rm  for}
\ \ \rho>\rho_0(\theta),
\end{equation}
 where the two constants do not depend on $\rho$ and $\theta$. Viceversa
  Eqs. \ref{support1}  and \ref{support2} impose that the support
of $\tau^{a\rho}$ lies in $
\rho<\rho_0(\theta)$.

In our formalism the metric assumes the form
\begin{equation}
\label{metric}
ds^2= (A_1^2-B_1^2) dt^2 +
2 (A_1 A_2-B_1 B_2)dtd\theta + (A_2^2-B^2_2) d\theta^2-d\rho^2.
\end{equation}
 while the determinant of the dreibein in polar coordinates is given by
\begin{equation}
{\rm det}(e)=A_2 B_1-A_1 B_2.
\end{equation}
Even though we shall in the following  be mainly interested in general case we
want to report what  happens in case of rotational symmetry. As derived in a
previous work \cite{MS2},
in the case of rotational symmetry, all functions, as expected,
do not depend on $\theta$. Furthermore from the two last symmetry equations
one obtains $\gamma_1=\gamma_2=0$, under the assumption that determinant never
vanishes. The regularity conditions at the origin for the functions $\alpha_i,\
\beta_i$ become
\begin{equation}
\label{regrot}
\alpha_1=O(\rho),~~~~\alpha_2=o(\rho^2),~~~~\beta_1=c+o(\rho),
{}~~~~\beta_2=1+O(\rho^2).
\end{equation}
and the only surviving symmetry equation is
\begin{equation}
\label{symmrot}
A_1\alpha_2-A_2 \alpha_1+B_2\beta_1-B_1 \beta_2=0.
\end{equation}
The support equations simplify to
\begin{equation}
\alpha_i^\prime=\beta_i^\prime=0\ \ \ \ {\rm and}\ \ \ \ \ \alpha_1\beta_2-
\alpha_2\beta_1=0
\end{equation}
outside the source. From these equations one can easily derive all
solutions  with rotational symmetry. (For more details see
ref. \cite{MS6}).

\section{Closed time-like curves and the weak energy condition}

For an arbitrary choice of the functions $\alpha_2$ and $\beta_2$ the
$g_{\theta\theta}=A^2_2-B^2_2$ term in the metric  is not
necessarily negative even  though, due to the regularity assumption,
$g_{\theta\theta}$ is negative in a neighbourhood of the origin. A not negative
$g_{\theta\theta}$ is a  symptom    of possible occurrence of CTC. In fact the
existence of CTC implies that $g_{\theta\theta}(\rho,\theta)$ is positive at
least for some $\rho$ and $\theta$. In fact given the CTC $t(\sigma )$,
$\rho(\sigma )$, $\theta(\sigma )$ at the point $\bar\sigma$, where
$t^\prime(\bar\sigma)=0$,  one would have $ds^2=g_{\theta\theta} d\theta^2
-d\rho^2 >0$.
For clearness sake we shall consider  first the case of rotational
invariance \cite{MS3,MS6}.
To begin with,
if the determinant of the dreibein  in the reduced radial gauge
vanishes at certain $\bar\rho$ it follows that the manifold at $\rho=\bar\rho$
either closes or become singular. Such a conclusion is obtained through the
following steps which are analyzed in detail in ref. (\ref{MS6}).
The regularity of the trace of the energy momentum tensor is an invariant
\begin{equation}
T^\mu_\mu= -{1\over \kappa}[{(\det (e))''\over \det (e)}+
{\alpha_1\beta_2 -\alpha_2\beta_1 \over \det (e)}]=-\frac{1}{2\kappa}~R.
\end{equation}
 On the other hand the term $\displaystyle{
\frac{(\alpha_1\beta_2-\alpha_2\beta_1)}{{\rm det}(e)}}$ is also an invariant
being the third eigenvalue of $T_{\mu\nu}$. Thus the regularity of the
remainder
imposes
\begin{equation}
\det(e)= c~ (\bar\rho-\rho) (1+O((\bar\rho-\rho)^2)).
\end{equation}
Now if in $\bar\rho$\ \  $A_2$ and/or $B_2\ne 0$ one can easily show that the
manifold is singular, while if $A_2=B_2=0$ in $\bar\rho$ the universe
closes without a singularity only if in $\bar \rho$ $A^2_1-B^2_1>0$ and
$\alpha_2^2-\beta_2^2=-1$. The topology of the resulting universe is that  of a
sphere and inside the universe ${\rm det}(e)\ge 0$.

If we now consider the WEC on the two light-like vectors $T^a+\Theta^a$ and
 $T^a-\Theta^a$ we obtain an inequality which is exactly integrable i.e.
\begin{equation}
\frac{d E^{(\pm)}}{d \rho} \le 0,
\end{equation}
where $E^{(\pm)} (\rho)\equiv
(B_2\pm A_2)(\alpha_1\pm\beta_1)-
(\alpha_2\pm\beta_2)(B_1\pm A_1)$. It is not difficult to show for a
conical universe, in absence of CTC  at infinity (which implies
$\alpha_2^2-\beta_2^2\le 0$), using ${\rm det}(e)>0$ and the support equation
$\alpha_1\beta_2-\alpha_2\beta_1=0$, that
$0\le E^{(\pm)}(\infty)\le E^{(\pm)}(\rho)$.
Then by straightforward algebra one obtains
\begin{equation}
\frac{d}{d\rho} \left (\frac{ A^2_2(\rho)-
B^2_2(\rho)}{{\rm det}(e)}\right )= -\frac{1}{2 {\rm det}(e)^2 }
[(A_2-B_2)^2 E^{(+)}(\rho)+(A_2+B_2)^2 E^{(-)}(\rho)]\le 0,
\end{equation}
and as $g_{\theta\theta}$ is negative at the origin it is always negative and
thus CTC cannot occur.
Such analysis can be extended to all universes with the single exception of the
cylindrical universe, generated by a string with tension and  zero angular
momentum.

All these reasoning can be also extended with no substantial change to the case
of an open universe not invariant under rotation, provided that in our
coordinates ${\rm det} (e)$ never vanishes \cite{MS7}.
Let us consider  in fact a general
external metric of the form
\begin{equation}
\label{metric1}
ds^2=g_{00}(\theta) dt^2  +2 g_{0\theta}(\rho,\theta) d t d \theta+
g_{\theta\theta}(\rho,\theta)  d \theta^2 - d \rho^2,
\end{equation}
i.e. the second order polynomial in $\rho$, \  $g_{00}(\rho,\theta)$  reduces
outside the source (or equivalently at  infinity) to a function of $\theta$.
We prove that $g_{00}(\theta)>0$ imposes that $\alpha_1=\beta_1=0$. In fact
such
a behaviour implies  $\alpha_1^2-\beta_1^2=0$ and $\alpha_1 A_1=\beta_1 B_1$.
Thus  if $\alpha_1\ne 0$ one has $A_1^2=B^2_1$ and $g_{00}\equiv 0$ (for
$\rho\ge \rho_0(\theta)$). Thus $\alpha_1=\beta_1=0$. Symmetry equations
(\ref{22a})  now
gives
\begin{equation}
(A_1+B_1)(\alpha_2-\beta_2)=-(A_1-B_1)(\alpha_2+\beta_2)
\end{equation}
and thus
\begin{equation}
(A_1^2-B_1^2)(\alpha_2^2-\beta_2^2)=g_{00}(\alpha_2^2-\beta_2^2)\le 0
\end{equation}
i.e. $\alpha_2^2-\beta_2^2\le 0$.
{}From $\alpha_1=\beta_1=0$ we have  the validity of the same support
equation
 $\alpha_2
\beta_1-\alpha_1\beta_2=0$ as in the
rotationally symmetric case. We are  thus
in the same situation as in the rotationally invariant case and thus we prove
that $g_{\theta\theta}(\rho,\theta )\le 0$. However CTC would imply that at
least for a value of $\rho$ and $\theta$   $g_{\theta\theta}(\rho,\theta )> 0
$ and thus there cannot be any CTC.

With regard to the metric it
is easy to prove \cite{MS7} that the assumption $g_{00}(\theta)>0$ implies that
$g_{0\theta}=g_{0\theta}(\theta)$ and thus the external metric assumes the form
\begin{equation}
\label{metric2}
ds^2=g_{00}(\theta)(d t+ J(\theta) d\theta)^2- (a(\theta)\rho-b(\theta))^2
 d\theta^2-d\rho^2
\end{equation}
because the coefficient $\gamma_{\theta\theta}$ of $d\theta^2$ is the square of
the dreibein determinant divided by $g_{00}$.
Performing the following change of variables
\begin{equation}
\theta^\prime=2\pi
\frac{\int^\theta_0 a(\phi) d\phi}{\int^{2\pi}_0 a(\phi) d\phi}
\end{equation}
and
\begin{equation}
t^\prime =t+\left [\int^{\theta(\theta^\prime)}_0 J(\phi) d\phi-
\frac{\theta^\prime}{2\pi}\int^{2\pi}_0 J(\phi) d\phi \right ]
\end{equation}
we reach the metric
\begin{equation}
\label{metric3}
ds^2=g_{00}(\theta)(d t^\prime + J_0 d\theta^\prime )^2-
(a_0\rho-b(\theta^\prime ))^2
 d\theta^{\prime 2}-d\rho^2
\end{equation}
where
\begin{equation}
a_0=\frac{1}{2\pi} \int^{2\pi}_0 a(\theta) d\theta
\end{equation}
and
\begin{equation}
J_0=\frac{1}{2\pi} \int^{2\pi}_0 J(\theta) d\theta.
\end{equation}
If $a_0\not =0$ one  easily proves that $g_{00}$ becomes a constant and we have
the usual conical metric \cite{flat}. If $a_0= 0$ we have a cylinder.

\section{Conclusions}

The application of the radial gauge to 2+1 dimensional gravity has
been successful both in dealing with extended sources and time
dependent problems. In ref. \cite{MS2,MS1} we gave the general resolvent
formulas for the time dependent problem in terms of a simple
quadrature and  derived  the support properties of the
energy momentum tensor in the case of time dependent sources with
rotational symmetry.
In ref. \cite{MS1}
we gave also explicit time dependent solutions, not necessarily
invariant under rotations which satisfy all energy conditions.
In the present paper we concentrated mainly on the general, non
rotationally invariant, stationary problem. We wrote down the
metric in terms of quadratures and gave explicit formulas for the
support of the energy momentum tensor. In addition we have shown
that the reduced radial gauge allows to derive important consequences
of the weak energy condition. In particular we proved that for the general
stationary open universe the WEC and the absence of CTC at infinity prevents
the occurrence of CTC everywhere, both in presence and in absence of
rotational symmetry.
The radial gauge approach appears also apt to examining the time
dependent situation in connection to 2+1 dimensional cosmology.

\end{document}